\documentclass{iaus}
\usepackage{graphicx}

\title[Subsurface flows and rotating sunspot] 
{Subsurface flows associated with rotating sunspots}

\author[Jain et al. ]   
{Kiran Jain, Rudolf Komm, Irene Gonz{\'a}lez Hern{\'a}ndez, Sushant C. Tripathy, and Frank Hill}

\affiliation{National Solar Observatory\\ 950 N. Cherry Av., Tucson, AZ 85719,USA \\email: {\tt kjain@noao.edu}}

\pubyear{2010}
\volume{273}  
\pagerange{119--126}
\jname{Physics of Sun and Star Spots }
\editors{D.P. Choudhary \& K.G. Strassmeier eds.}
\begin{document}

\maketitle

\begin{abstract}
 In this  paper, we compare components of the horizontal flow  below the solar 
surface in  and around  
regions consisting of rotating and non-rotating sunspots. 
 Our  analysis suggests that there is 
a significant variation in both components of the horizontal flow  
at the beginning  of sunspot rotation as compared to  the non-rotating 
sunspot. The flows in surrounding areas are in most cases relatively small.
  However, there is a significant influence of the motion on flows in an area closest
 to the sunspot rotation.
\keywords{Sun : helioseismology, Sun : interior, (Sun) : sunspots}
\end{abstract}

\firstsection 

\section{Introduction}
Sunspots that rotate around their umbral centers are termed as rotating sunspots. In some cases, these are identified by the rotation around another sunspot within the same active region (AR). 
 The identification of these sunspots has been made easier by advances in the spatial 
and temporal resolution of recent satellite-borne telescopes (e.g. \cite[Brown \etal\ 2003]{Brown03}). The origin of rotational motion is believed to be due to the shear and twist in magnetic field lines or vice-versa. It is also suggested that the magnetic twist may result from large-scale flows in the solar convection zone and the photosphere or in sub-photospheric layers. In an earlier study,  \cite[Zhao \& Kosovichev (2003)]{zhao03} found evidence for two opposite sub-photospheric vortical flows in the depth range of 0--12 Mm around a fast rotating sunspot in AR 9114.
In this paper, we extend our earlier study of the flows beneath a rotating sunspot (\cite[Jain \etal\ 2010a]{jain10a}) to the surrounding regions,  and investigate how flow fields change with depth compared to a non-rotating case.

\section{Technique }
We apply the technique of ring diagrams to obtain the depth dependence beneath the solar surface (\cite[Hill 1988]{hill88}).
In this technique, the high-degree waves in localized areas over the solar surface are used to infer the 
characteristics of these propagating waves.  We use high-cadence continuous Dopplergrams obtained by the 
 Global Oscillation Network Group (GONG).
These images are processed through the GONG ring-diagram pipeline 
(\cite[Corbard \etal\ 2003]{corbard03})
where we track and remap each region  for 1680 min, and apply a    three-dimensional FFT  
on both 
spatial and temporal direction to obtain a 3D power spectrum. The corresponding power spectrum is fitted using a Lorentzian profile
model (\cite[Haber \etal\ 2000]{Haber00}) that includes the perturbation term due to horizontal
flow fields in the region.  Finally, the obtained velocities are inverted using regularized least square 
method to obtain the depth dependence of the horizontal velocity flows. 

\section{Data and Analysis}

We study flows in photospheric layers beneath regions  in and around AR 10930  and AR 10953.
AR 10930 was observed during the declining phase of solar cycle 23 and located at E19S04
 on 2006 December 10.  It had two major sunspots; the big sunspot
did not show any visible change during the period of disk passage  while the small sunspot in the southern part exhibited rapid counterclockwise rotation about its umbral center. Since we do not have the resolution to discriminate between
two sunspots, for our purpose, we will consider the group as a ``rotating sunspot".
The sunspot started to
rotate after mid day on December 10 and continued to rotate until December 13.  
Since the ring-diagram inferences are affected by the location and size of the region,we  compare flows in the region of rotating sunspot  with those for an active region with non-rotating sunspots located at around the same location as AR 10930. 
For this purpose, we analyze AR 10953 that was located at S10E09 on 2007 May 1 with non-rotating sunspots. Further,  both active regions are approximately of the same size.We use three consecutive time series to study $x$- and $y$-components of the horizontal flow; these are chosen in such a way that they represent the period of before, during and after the beginning of sunspot rotation in AR 10930.  For
comparison, we also use three time series of equal length for AR 10953.

We calculate flows in a mosaic of 9 regions where each region is   $\sim$ 11$^{\mathrm{o}}\times$11$^{\mathrm{o}}$  in size. The central region  in this mosaic has the active region at its center and is surrounded by eight regions spaced by 5$^{\mathrm{o}}$ in each direction.
The depth variations of $x$- and $y$-components of horizontal velocity in the mosaic of  AR  10930 for all three time samples are shown in Figures~1 and 2. Note that the active
region is located at the center of Panel ({\it e}). Since AR 10930 was a medium sized active region, the central patch covers an area  larger than the actual sunspot group. It is clearly seen that the relative values of $u_x$ and $u_y$ in Panel ({\it e}) of  both figures are significantly larger than in the neighboring regions which are relatively
quiet. This is in agreement with earlier findings where flows in active regions are larger than their quieter surroundings (\cite[Komm \etal\ 2005]{komm05}).

\begin{figure*}[t]
\begin{center}
\hspace{3.5pc}
\includegraphics[width=70mm,angle=90]{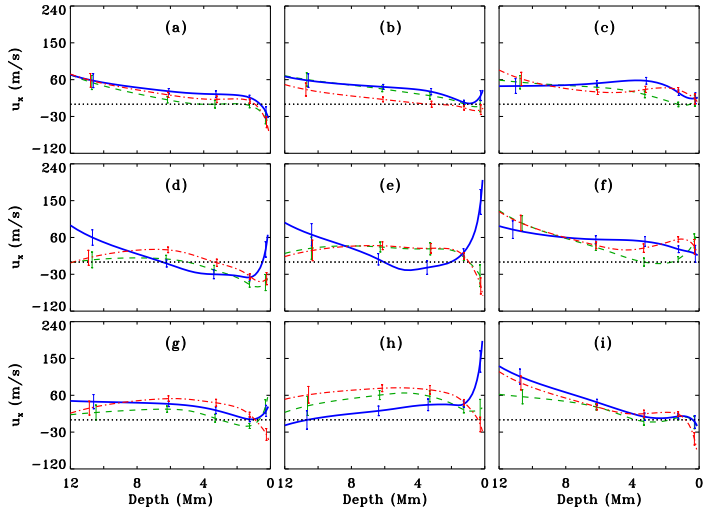}
\caption{Depth variation of $x$-component of horizontal flows in and around AR10930 for three overlapping time series; green/dashed line represents the period before the beginning of rotation,
blue/solid line includes the period when the rotation started, and red/dash-dot line is for
after the rotation began.  Error bars are shown for a few points. The active region is located at the center of Panel ({\it e}).   }
\label{label1}
\end{center}
\end{figure*}
\begin{figure*}[h]
\begin{center}
\hspace{3.5pc}
\includegraphics[width=70mm,angle=90]{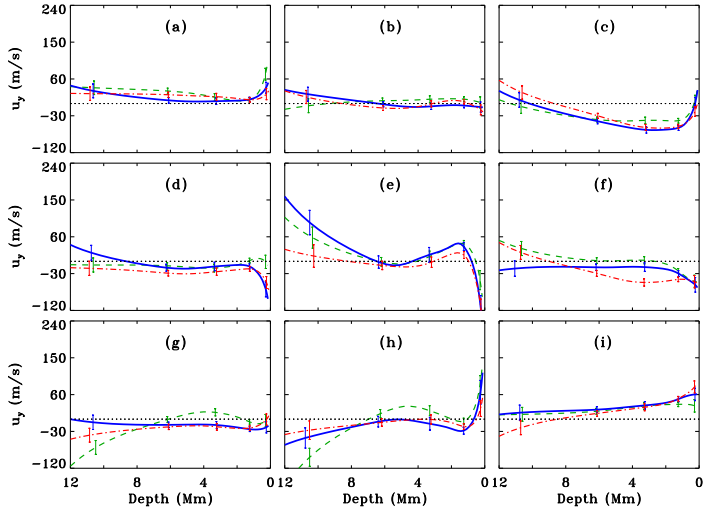}
\caption{Same as Figure~1 but for the $y$-component of the horizontal flow.}
\label{label2}
\end{center}
\end{figure*}

It is evident from Figure~1({\it e}) that the  maximum variation in $u_x$  with depth is obtained at the beginning of sunspot rotation while no  
significant variation is seen on the other two days. 
Here  we focus on the depth range between 3 Mm and 12 Mm, where the errors are
 smaller due to the set of modes contained in the analysis combined with the 
limitations of the model used near the surface. Below 3 Mm, we find that 
 $u_x$, at the beginning of sunspot rotation,  increases rapidly in deeper layers. 
Although $u_x$ has significantly decreased in surrounding regions, 
there is a significant influence of
rotation on the profiles of $u_x$ in the regions adjacent to the rotating part
of the sunspot group.  The values of $u_x$ in Panels ({\it d}) and ({\it h}) 
are again larger at the beginning of sunspot rotation while profiles for all three cases 
 did not change significantly in other neighboring regions. 

On the other hand, even if profiles of $u_y$ in Panel ({\it e}) of Figure~2 are similar for all three time samples, these exhibit variations in magnitude.  
It starts increasing below 6 Mm and the maximum value is again 
achieved at the beginning of sunspot rotation. 
Since the region is located in the southern hemisphere and 
the $u_y$ shows a strong poleward trend, this
 significantly higher positive value of $u_y$ during the 
initial stages of  sunspot rotation indicates an equator-ward meridional flow. 
However, neighboring regions in this case
are less influenced by the sunspot rotation. In contrast to the rotating sunspot region, the flows in a non-rotating sunspot region show less variation in both components. As an example, we exhibit the depth variation  of $u_x$ and $u_y$ beneath AR 10953 in Figure~3. The profiles for both components are similar for all three time samples. We also find that flows  in surrounding area are comparable for all three time samples but with smaller magnitude.

\section{Summary}
We have compared the horizontal flow components in and around two
sample cases of rotating and non-rotating sunspots. Our analysis
suggests that both $u_x$ and $u_y$ show significant variation with depth during 
the course of sunspot rotation as compared to the non-rotating sunspot. 
The maximum change in $u_x$ is found at the beginning of sunspot rotation.  
The flows in surrounding regions in both cases are relatively small, however there is a significant variation in regions neighboring to the rotating sunspot.
In the present study, we have considered a region that includes both
 rotating and non-rotating sunspots, and the quiet area around them. 
This is mainly due to the limitation imposed by the resolution of images 
on the technique.   However, the high-resolution images from Helioseismic 
Magnetic Imager (HMI) onboard {\it Solar Dynamics Observatory (SDO)}  will allow 
us to study smaller regions within an active region that will provide 
a deeper insight on the dynamics of subsurface
layers beneath individual sunspots, e.g. using HMI test Dopplergrams, \cite[Jain \etal\ (2010b)]{jain10b} have shown that the ring-diagram technique can be reliably applied to regions as small as 5$^{\mathrm{o}}$.
\acknowledgements
This work utilizes data obtained by the Global Oscillation Network
Group (GONG) project, managed by the National Solar Observatory, which
is operated by AURA, Inc. under a cooperative agreement with the
National Science Foundation. The data were acquired by instruments
operated by the Big Bear Solar Observatory, High Altitude Observatory,
Learmonth Solar Observatory, Udaipur Solar Observatory, Instituto de
Astrof\'{\i}sica de Canarias, and Cerro Tololo Interamerican
Observatory. This work was supported by NASA-GI grant NNG 08EI54I.
\begin{figure*}
\begin{center}
\includegraphics[width=60mm]{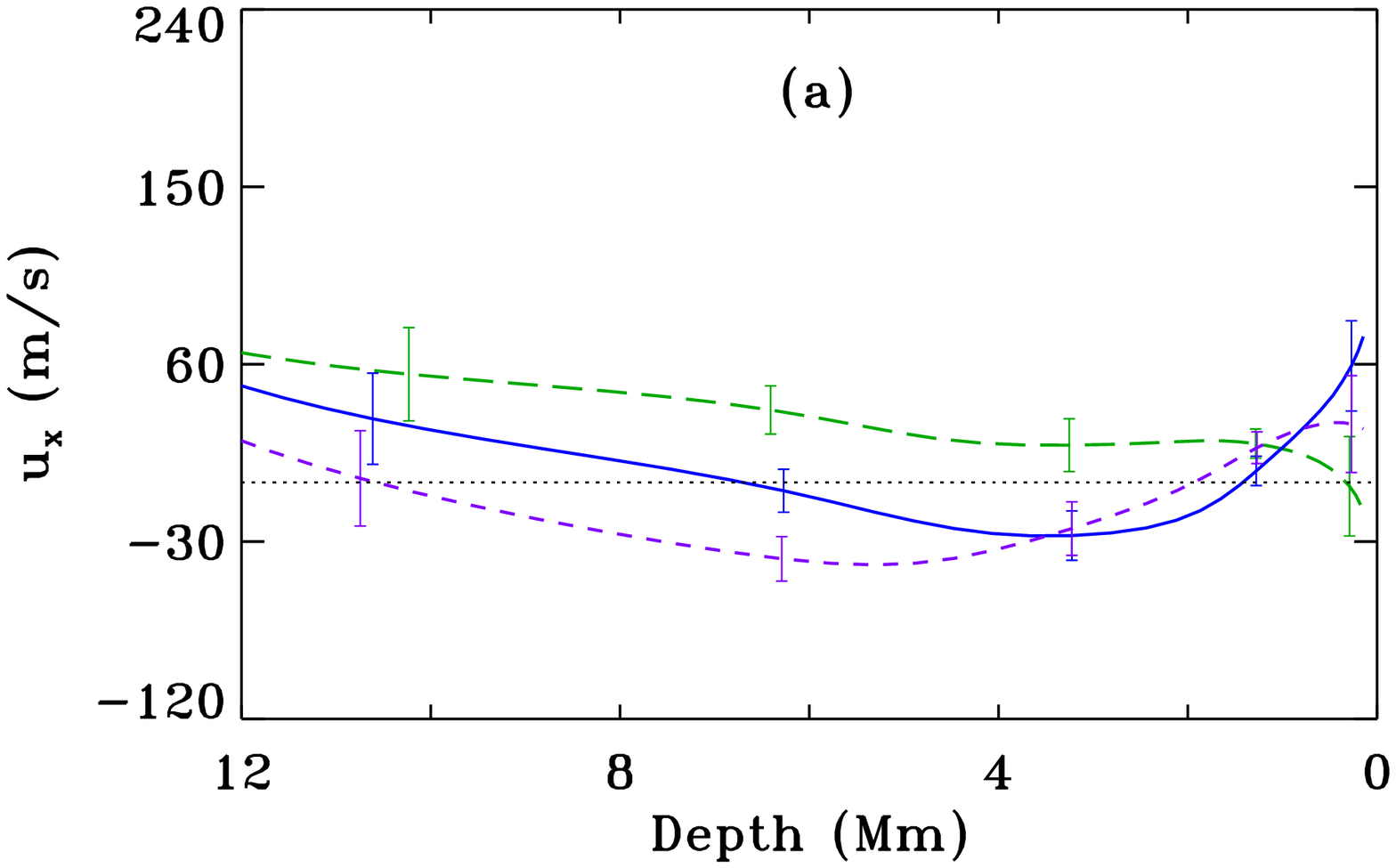}
\includegraphics[width=60mm]{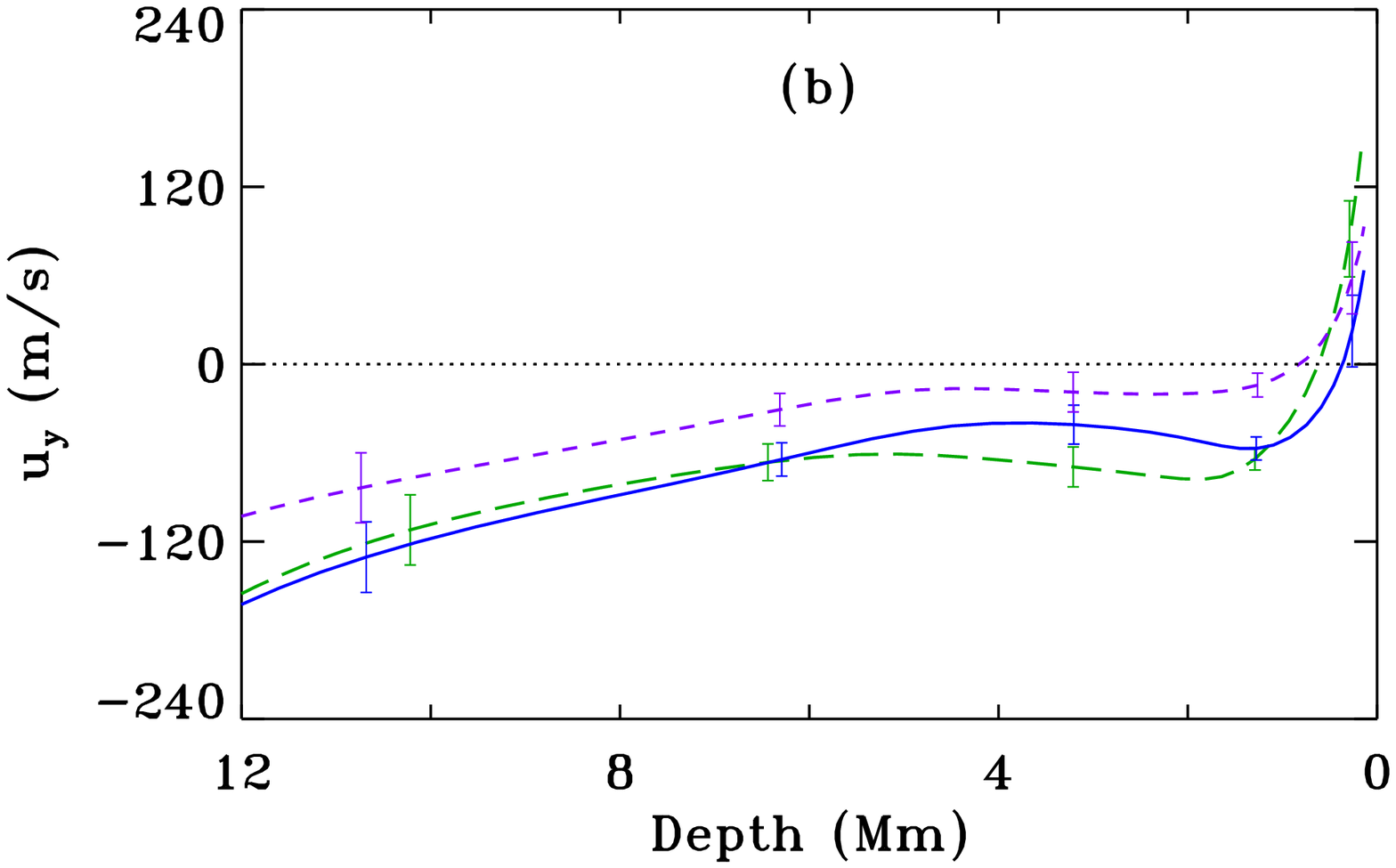}
\caption{Depth variation of ({\it a}) $x$- and, ({\it b}) $y$-components of the horizontal velocity flow for AR10953  for three overlapping time series; green/long-dash, blue/solid, and  purple/small-dash are for time samples 1, 2 and 3 respectively. The errors are shown only at selected depths.}
\label{label6}
\end{center}
\end{figure*}

\end{document}